\documentstyle[11pt,aaspp4,epsf]{article}
\slugcomment{Accepted for publication in The Astrophysical Journal}
\begin{document}

\title{The roughness of the last scattering surface}
\author{Silvia Mollerach$^{1,5}$, Vicent J. Mart\'\i nez$^1$,
J. M. Diego$^{2,3}$, E. Mart\'\i nez-Gonz\'alez$^2$, 
J. L. Sanz$^2$, and Silvestre Paredes$^4$}
\affil{$^1$ Departamento de Astronomia y Astrof\'\i sica, 
Universidad de Valencia, E--46100 Burjasot, Valencia, Spain\\
$^2$ IFCA, Departamento de Astrof\'\i sica, Facultad de Ciencias, 
Avda. Los Castros, s/n, Santander 39005, Spain\\
$^3$ Departamento de F\'\i sica Moderna, Facultad de Ciencias, 
Avda. Los Castros, s/n, Santander 39005, Spain\\
$^4$ Departamento de Matem\'atica Aplicada. Universidad de Murcia.
Paseo Alfonso XIII, E--30203 Cartagena, Murcia, Spain\\
$^5$ Departamento de F\'\i sica, Universidad Nacional de La Plata,
cc 67, La Plata 1900, Argentina}
\begin{abstract}
We propose an alternative analysis of the microwave background  
temperature anisotropy maps that is based on the study of the 
roughness of natural surfaces. We apply it to large angle 
anisotropies, such as those measured by {\it COBE}--DMR. 
We show that for a large signal to noise 
experiment, the spectral index can be determined independently of 
the normalization. We then analyze the 4 yr {\it COBE} map and find
for a flat $\Omega=1$ universe, 
that  the best-fitting value for the spectral index is
$n = 1.15^{+0.39}_{-0.34}$
and for the amplitude $Q_{rms-PS}= 14.1^{+3.9}_{-3.5}\mu K$. 
For $n=1$, the best-fitting normalization is $Q_{rms-PS}|_{n=1}= 
16.2^{+1.4}_{-1.3}\mu K$.
\end{abstract}

\singlespace
\section{Introduction}

The measurement of the cosmic microwave background  
temperature anisotropies at large angular scales by the {\it COBE}
satellite has provided a unique insight to the universe 
inhomogeneities at the recombination time. It is believed that
these anisotropies originated from small perturbations in
the gravitational potential and thus they give information 
about the  inhomogeneities at the linear evolution stage, 
avoiding most of the complications associated to
the study of the non-linear evolution of structures.

Since the discovery of the anisotropies, several statistical 
analysis of the data were proposed. The most widely used ones are
the two point correlation function (\cite{hi96a}) and the angular
power spectrum (\cite{wr96,go96,hi96b,te96,te96,ba96,bu97,bo97}). 
These statistics contain a complete description of the anisotropies
if they are Gaussian distributed, and thus are the most natural
choice to test cosmological models leading to Gaussian primordial
fluctuations. However, there are also possible physical mechanisms
which produce non--Gaussian distributed anisotropies. Motivated by 
this possibility, other statistical analysis of the anisotropy 
maps, which are not shaped for Gaussian fields, have been developed. 
These include the study of higher order correlation 
functions (\cite{hi95}) and 
morphological characteristics of hot and cold spots, 
as genus and density of spots (\cite{to95,ko96,fa96}), 
Minkowski functionals (\cite{sc98}) and
multifractal studies (\cite{po95,di99}). 
It is quite remarkable that \cite{go95} find a fractal behavior for 
the {\it COBE}--DMR first year data using the perimeter--area 
relation of iso--temperature contour regions. Whereas the topological
analyses give information about the invariant quantities under
deformations, the multifractal analyses shed light on the  
self-similarity of the field.
Although the first analysis of the {\it COBE} maps, based on the 
three-point correlation function indicated that the anisotropies
are consistent with Gaussianity (\cite{ko96}), the recent analysis 
of the bispectrum by \cite{fe98} argues for evidence of 
non--Gaussianity in the {\it COBE} maps. Whether this effect can be
explained by some still unrecognized foreground contamination
or is a real signal of CMB non--Gaussianity is still an open question.

Both sets of statistical analysis, those based on the angular spectrum
and the topological ones, have been extensively applied to the 
{\it COBE}--DMR anisotropy maps and several determinations of the 
amplitude of the inhomogeneities (usually parameterized in terms 
of the resulting quadrupole amplitude $Q_{rms-PS}$, that we 
abbreviate as $Q$) and of the primordial spectral index $n$ 
have been obtained. These determinations had an extraordinary 
impact on cosmological models as they provide the first accurate 
measure of the amplitude of the primordial fluctuations
(see \cite{be96,ba97} for a summary of the 4 yr {\it COBE} analysis and
the references above for the other analyses). Most of these
determinations show a strong anticorrelation between the 
inferred values for $Q$ and $n$.

\begin{figure} 
\begin{center}
\epsfxsize=6.5cm
\begin{minipage}{\epsfxsize}\epsffile{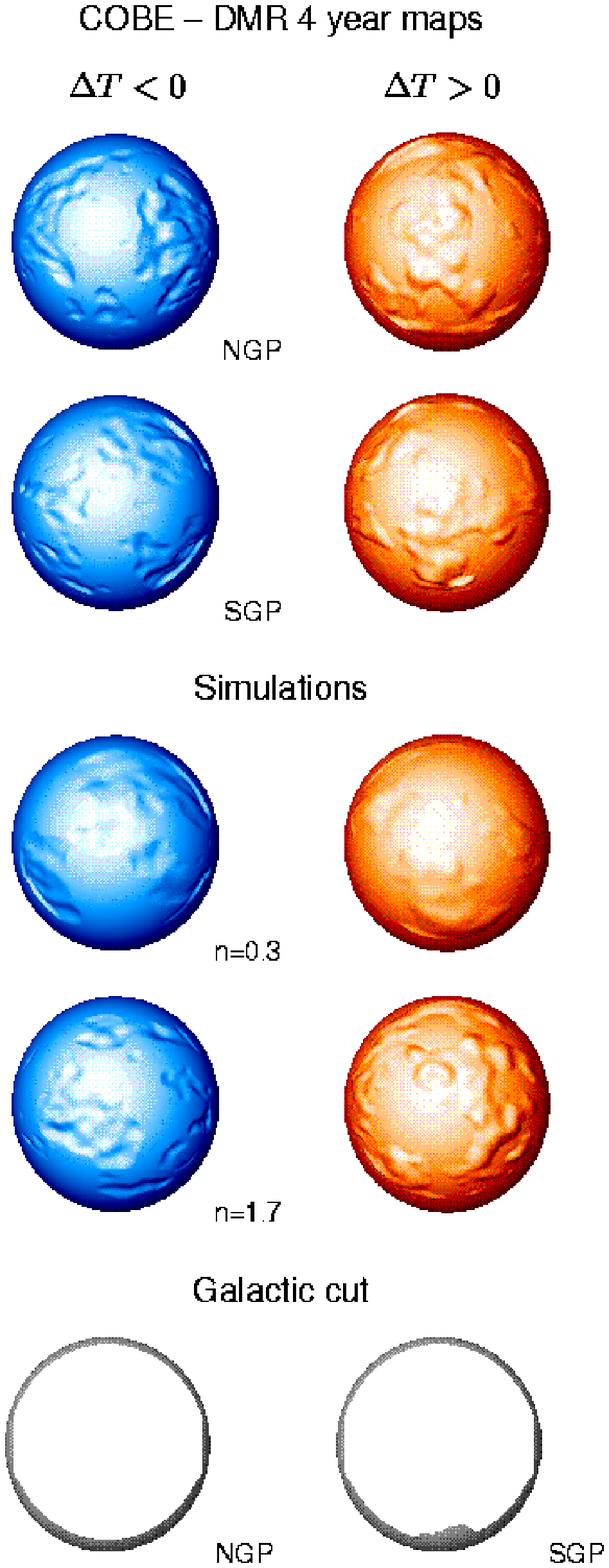}\end{minipage}
\caption{ The four top spheres represent the 
anisotropies of the 4--year {\it COBE}--DMR maps. 
The two upper ones are centered in the North Galactic Pole (NGP), 
while the other two are in the South Galactic
Pole (SGP). On the left--hand side, in blue,  
pixels where $\Delta T <0$ 
are depicted as depths, while on the right--hand side, in red, 
pixels where $\Delta T >0$, are displayed as heights. 
The `sea--level' in each blue (red) sphere corresponds to 
the pixels where  $\Delta T \ge 0$ ($\Delta T \le 0$). 
Below, spheres centered only at the NGP are similarly displayed and
correspond to two simulations with spectral indices $n=0.3$ 
and $n=1.7$. A smoothing filter of $7^\circ$
(full width at half maximum) has been applied to the {\it COBE}
data to reduce the effect of the radiometer noise.
In all cases the height of the pixels lying within the 
customized Galactic cut has been fixed to zero, and therefore
those pixels belong to the `sea--level'. The segments of the 
maps corresponding to the Galactic cut are displayed in grey 
at the bottom panels.}
\end{center}
\label{spheres}
\end{figure}

We propose here an alternative approach to the analysis of the 
anisotropies, which is inspired in the study of the 
roughness of natural surfaces in fields like geology, biology
or metallurgy. Many of the natural surfaces can be modeled
as fractals and a central concept related to their roughness
is that of the fractal dimension. The dimension index for
fractal surfaces lies between 2 and 3 and it depends on the 
irregularity of the surface. By associating to every point 
in a sphere a height (or depth) proportional to the excess 
(or defect) of the temperature measured in that direction with 
respect to the mean, we obtain an irregular surface which 
roughness is directly related to the temperature anisotropies.
In Figure \ref{spheres} we show the {\it COBE}--DMR
4 yr map of the CMB anisotropies by means of this procedure.
The blue (red) spheres represent the pixels where $\Delta T <0$
($\Delta T >0$) as depths (heights) over a `sea--level'.
In each pair of complementary spheres the `sea--level'
represents the pixels where $\Delta T \ge 0$ for the blue ones 
and $\Delta T \le 0$ for the red ones.
In this paper, we study how to obtain information about the
primordial inhomogeneities from the analysis of the roughness
of these surfaces using fractal techniques.

The outline of the paper is as follows. In section 2 we 
briefly describe the most appropriate approach to measure the 
surface fractal dimension, we present an analytical relation 
between the spectral index and the fractal dimension and discuss 
the limitations of the procedure of trying to measure the 
fractal dimension and to determine 
the spectral index for actual sky data. In section 3 we apply 
these techniques to simulations of the {\it COBE}--DMR maps, and 
present a method to determine the spectral index, when the 
radiometer noise can be neglected. In Section 4 we present
a modified method which can be used in small signal to noise maps to
obtain a joint determination of $Q$ and $n$, we test it with 
simulations of the {\it COBE}--DMR maps, and we apply it to 4 yr 
map. Section 5 contains the conclusions.

\section{Determination of the fractal dimension of a surface}

Simple objects like points, segments, disks or spheres have
dimensions 0, 1, 2 and 3 respectively. In between them there
exist more complex objects with non-integer dimension. Irregular
surfaces have fractal dimension between 2 and 3 depending on how
much volume they fill. The fractal dimension is thus a useful
tool to describe the roughness of surfaces (\cite{ma82}). 
Different methods 
have been proposed in the literature to estimate it. 
One possibility could be to section the surface with 
a plane, measure the dimension of the profile and use the fact 
that the fractal dimension of the surface is one plus the dimension
of the profile. The other possibility is to use algorithms
that directly estimate the dimension of the surface. Some of 
the algorithms to evaluate the fractal dimension of profiles 
and surfaces have been comparatively tested by \cite{du89a} and
\cite{du89b}, who also proposed an improved algorithm, the 
variation method. This is specially well suited when 
dealing with profiles or surfaces defined by a function of one
or two variables respectively. In these cases one of the 
coordinates describing the profile or surface is of a 
different nature than the other(s).
For example, the surface that we want to analyze 
can be described by two angular coordinates and a third one 
measuring temperature differences. 
One would like that the associated dimension be independent of the 
choice of the units in which the function is measured. 
The variation algorithm assigns the same
fractal dimension to the surfaces defined by a function $f$ and by
the function $Cf$, with $C$ a constant (it is invariant under 
affine transformations). 
The various methods to estimate the fractal dimension are obtained
from the discretization of different definitions. Although they are
mathematically equivalent in the continuum limit (\cite{fa90}),
they can give different 
results when adapted to  discrete sampled data. This is the case for 
the Minkowski--Bouligand (MB) dimension and the box dimension.
They can both be defined as
\begin{equation}
D=\lim_{L \to 0} \left(3-\frac{\log V(L)}
{\log L}\right),
\label{df}
\end{equation}
where $V(L)$ is the volume of a covering of the surface, 
which in the case of the MB dimension is the set of all points
at a distance less than $L$ from a point of the surface 
(or the union of all the balls of radius $L$ centered 
on a point of the surface). In the box dimension case, 
$V(L)$ is the volume of the union of disjoint cubes
of side $L$ needed to cover the surface.
Algorithms implementing these definitions calculate $V(L)$
for some values of $L$ and then recover $D$ from the slope of the
$\log V$--$\log L$ plot. Both methods have the problem pointed 
out before,
they give different results when applied to functions $f$ and
$Cf$. The variation method (\cite{du89b}) overcomes 
this difficulty by defining a new covering  of the surface 
in the following way. If  the surface is described as a set of points
$(x,y,f(x,y))$, it associates to each point of the surface 
an upper point as the maximum of the function in an 
$L$--neighborhood $[x-L,x+L] \times 
[y-L,y+L]$ and a lower point as the minimum of 
the function in that neighborhood. $V(L)$ is the volume 
of the region between the upper and lower surfaces
(or the volume of the union of all the squares $[x-L,x+
L] \times [y-L,y+L]$ centered on a 
point of the surface and truncated to the region $(x,y)$
covered by the surface). Once again, $D$ is obtained from the
slope of the $\log V$--$\log L$ plot. Theoretically, we would 
expect a constant slope for a fractal surface. However, 
the constant slope is only obtained in practice for an 
intermediate range of scales, which
are large enough so that there is a statistically good
sampling of every point neighborhood, and small enough 
compared to the total sample size. As we will see, these 
conditions are problematic for the application to 
large angle anisotropy maps, as the DMR ones. Nevertheless, 
the method can be adapted to extract useful information from 
the data.

To illustrate the connexion between fractals and the CMB, let us
suppose that the Sachs Wolfe effect gives the main contribution to the
anisotropies. In this case, large angle CMB anisotropies can be described 
as a fractal, with dimension related to the spectral index 
of the gravitational potential from which they originated.
They resemble in fact one  of the most frequently used
mathematical models to construct random fractals, the 
fractional Brownian functions (FBF) (\cite{ma86}).
For one variable, the fractional Brownian motion is a
single--valued function, which increments, $F(t_2)-F(t_1)$,
have a Gaussian distribution with variance
$\langle |F(t_2)-F(t_1)|^2 \rangle \propto 
|t_2-t_1|^{2 H}$,
where the brackets denote the ensemble average over many 
samples, and the parameter $H$ has a value $0<H<1$. The 
value $H=1/2$ corresponds to the well--known Brownian motion.
The FBF show a statistical scaling behavior: for a function
of $E$ Euclidean dimensions, if ${\bf x}=(x_1,...,x_E)$ is scaled 
by a factor $r$, $\Delta F$ is
changed by a factor $r^H$, namely
$\langle |\Delta F(r {\bf x})|^2 \rangle = r^{2 H} 
\langle |\Delta F({\bf x})|^2 \rangle$.
This scaling by a different factor for ${\bf x}$ and $F({\bf x})$ 
is known as self--affinity. It can be seen that the fractal dimension 
associated to this kind of function is $D=E+1-H$ (\cite{vo88}).
 The power spectrum of a FBF is a 
power law and the spectral index is directly related to the 
parameter $H$, and thus to  $D$. 
In fact, a widely used algorithm to construct a 
FBF with a given fractal dimension $D$ is to obtain a random 
function with the required power--law spectrum
by a Fourier transformation (\cite{sa88}). 
According to this,
the CMB anisotropies originated through the Sachs--Wolfe effect
by a gravitational potential $\Phi({\bf x},t)$ with power 
spectrum $P(k)=A k^{n-4}$ in a flat $\Omega=1$ universe
can be described by a FBF. 
We can directly obtain the parameter $H$
from the computation of
\begin{equation}
\langle |\Phi({\bf x}+{\bf y})- \Phi({\bf x})|^2 \rangle=
\frac{A}{\pi^2}\int_0^\infty
dk\ k^{n-2}\left(1-\frac{\sin(ky)}{ky}\right)
\propto y^{1-n}\label{fbm}
\end{equation}
where $y$ is the physical distance between the two points.
If we consider points on a sphere, it is given by 
$y=2 r \sin(\theta/2)$, with $r$ the radius of the sphere and 
$\theta$ the angular separation between the points.
This relation holds for $-1<n<1$, that is the range in which 
the integral in eq. (\ref{fbm}) converges. Thus, $H$
is directly related to the spectral index $n$ by the equality
$2H=1-n$. We can hence in principle determine the spectral 
index $n$ by measuring the fractal dimension of the surface 
constructed by assigning to every point on a sphere a height
(or depth) proportional to the temperature excess (or defect)
in that direction. These are related by $D=3-H=(5+n)/2$.
Spectral indices larger than $n=1$ saturate in the fractal dimension 
$D=3$ as a surface cannot have a larger fractal dimension.
We would like to stress that the fractal dimension is
independent of the  spectrum normalization, 
it only depends on the spectral index. This is due 
to the fact that $D$ does not change when 
the function is multiplied by a constant factor because of the 
self--affinity property already discussed. 
Let  us note that this discussion applies to the simplest
case of a flat universe, with vanishing cosmological constant
and constant spectral index at the large angular scales
tested by COBE, where the Sachs--Wolfe effect gives the 
dominant contribution to the anisotropies. In the case of a
non-vanishing cosmological constant or an open universe there are
additional contributions to the anisotropies due to the time variation
of the gravitational potential along the photons path. At smaller
angular scales the acoustic oscillations of the photon-baryon plasma
at recombination give the dominant contribution to the anisotropies
and thus no scaling is expected.

In practice, however, it is not possible to determine $n$
through the above mentioned analytical relation with $D$. There are some 
difficulties to use it in actual maps of the sky measuring
large angular scale anisotropies as the DMR ones. The main
problem comes from the fact that the size of the surface is 
limited by the solid angle of the sphere (in fact, even smaller
due to the Galactic cut). This constrains the maximum number 
of points at which the surface is sampled.
The {\it COBE}--DMR maps give the temperature at 6144 pixels, that 
reduce to approximately 4000 when the Galactic cut is taken 
into account. 
In order to be able to determine the fractal 
dimension $D$ of a surface from the slope of the 
log$V$--log $L$ plot, 
we need to have a large enough sampling of the 
surface so that we pick a range of $L$ values in which 
the plot has a constant slope (we have already pointed out 
that deviations of this behavior are expected both at large 
scales, close to the size of the sample, and small scales,
close to the spacing of the sampling points). We have found
using synthetic fractals that in order to see the constant slope
region in the plot we need a sampling of at least $10^6$ points. 
We constructed two-dimensional FBF with various values of the 
fractal dimension and for various sizes of the sample. Only 
when the number of sampling points was larger than $(1024)^2$ 
a plateau in the $D(L)$ plot was evident.
This means that we would need to have a much larger
number of pixels than in the {\it COBE} maps. If we have to work 
with a smaller number of pixels, the slope of the plot will 
not show a plateau and we cannot obtain $D$ directly from its 
value (and $n$ from the analytical relation). 
Nevertheless, even if the slope varies 
with the scale, surfaces constructed with different $n$ 
 show a different behavior of the slope vs. scale 
relation, $D(L)$, and this can be used to identify the underlying 
spectral index, as is shown in the next section.
Another point to be taken into account is that the maps
 have the dipole subtracted. The dipole is mostly due to
the proper motion of the Earth, and as this contribution cannot
be separated from a primordial one, it is completely subtracted 
from the maps. This procedure breaks the scaling present in eq.
(\ref{fbm}) and hence also interferes in the determination of $D$. 
The last point that we have to consider is the effect of 
the instrumental noise, but we will leave its discussion 
until section 4. Let us first consider the determination 
of $n$ in a very high signal to noise experiment.

\section{Ideal noiseless experiment}

We can see that the slope of the $\log V$--$\log L$
plot has a strong enough dependence on the spectral index to
be a useful tool to determine $n$ by
testing the predictions of a large number of Monte Carlo
simulations of the  maps for a range of $n$ values (we assume a flat
$\Omega=1$ universe).
For definiteness, we consider the {\it COBE}--DMR experiment with 
all its  actual characteristics, as pixelization and beam smearing 
(\cite{be96}), but we neglect for the moment the radiometer
noise. We use a $|b| < 20^\circ$ Galactic cut.
Thus, considering the quadrilaterized spherical cube 
pixelization used by the {\it COBE} team and cutting the Galaxy, 
our surface consists of two pieces corresponding 
to the two uncut faces of the cube, including the north and
south Galactic poles respectively, with adjacent strips 
coming from the cut faces. Each complete face has
$32^2$ pixels, while the strips have approximately 8 pixels
width. For each of these pixels, we have a value of $\Delta
T/T$ as it would be measured by an observer with an ideal
radiometer.

We have adapted the variation method (\cite{du89b})
to this surface in order to determine the  
$V(L)$ curve. In this algorithm, the 
volume associated to a scale $L$ is given by  
the region located between a top and a bottom surface 
constructed in the following way. To each pixel we associate an
upper (lower) point at scale $L$ by the maximum
(minimum) of the $\Delta T/T$ corresponding to the pixels in 
a square of side $L$, centered on that pixel. These
points form the top (bottom) surface. We only consider scales
corresponding to an integer number of pixels. 
Points located near the 
borders, i.e. close to the Galactic cut are problematic.
Standard border corrections (as multiplying by the fraction 
of the area corresponding to the missing region) are not 
useful to look for maxima and minima, thus we deal with 
the problem in another way. First, we consider
the surface formed by the two complete faces,
the north and the south ones, and use the pixels in the strips
coming from the cut faces only for the computation of the 
maxima and minima to be associated to the points close to 
the borders in the north and south faces. This is a compromise 
solution between loosing some information from the cut
face pixels and reducing the need for border 
corrections. At the largest scales, where problems with 
edge effects remain for some pixels, 
we changed the region to another one,
with approximately the same area in the pixel neighborhood.
We followed the implementation of the algorithm proposed  
by \cite{du89b}. After obtaining the volume associated to 
the surface at scales  $L$
from 2 to 30 pixels (in steps of 2), we computed the local dimension
as the slope of a five-point log-log linear regression of 
$V/L^3$ vs. $1/L$. This local dimension is regarded as
a `sliding window' estimate (through the scale) of the
fractal dimension (\cite{du89a}). As discussed above,
a constant slope $D$ characterizes a fractal surface of fractal
dimension $D$.

We also perform a second kind of analysis to the maps, that
makes full use of the cut faces data and does not suffer
from border effects. This consists in making a one--dimensional
study of the $\Delta T/T$ profiles along lines parallel to the 
Galactic cut. As these are closed lines, there are no borders.
There are seven of these complete lines of pixels to the north 
and other seven to the south of the Galactic cut, each one is 
128 pixels long. The analysis of these profiles is similar to that
performed on the surface, but now we compute the area of the
surface covering the profile. 
For each of the 14 profiles that we analyze in a
map, we obtain the area $S$ of the covering surface at 
scales $L$ from 3 to 41 (in steps of 4) and we take a mean 
over them. We then compute the slope of the 
$\log (S/L^2)$ vs. $\log (1/L)$ plot by fitting a line to each
group of five consecutive points. A fractal profile with 
dimension $D$ (between 1 and 2) would have a constant slope
equal to $D$.

\begin{figure}
\epsfbox{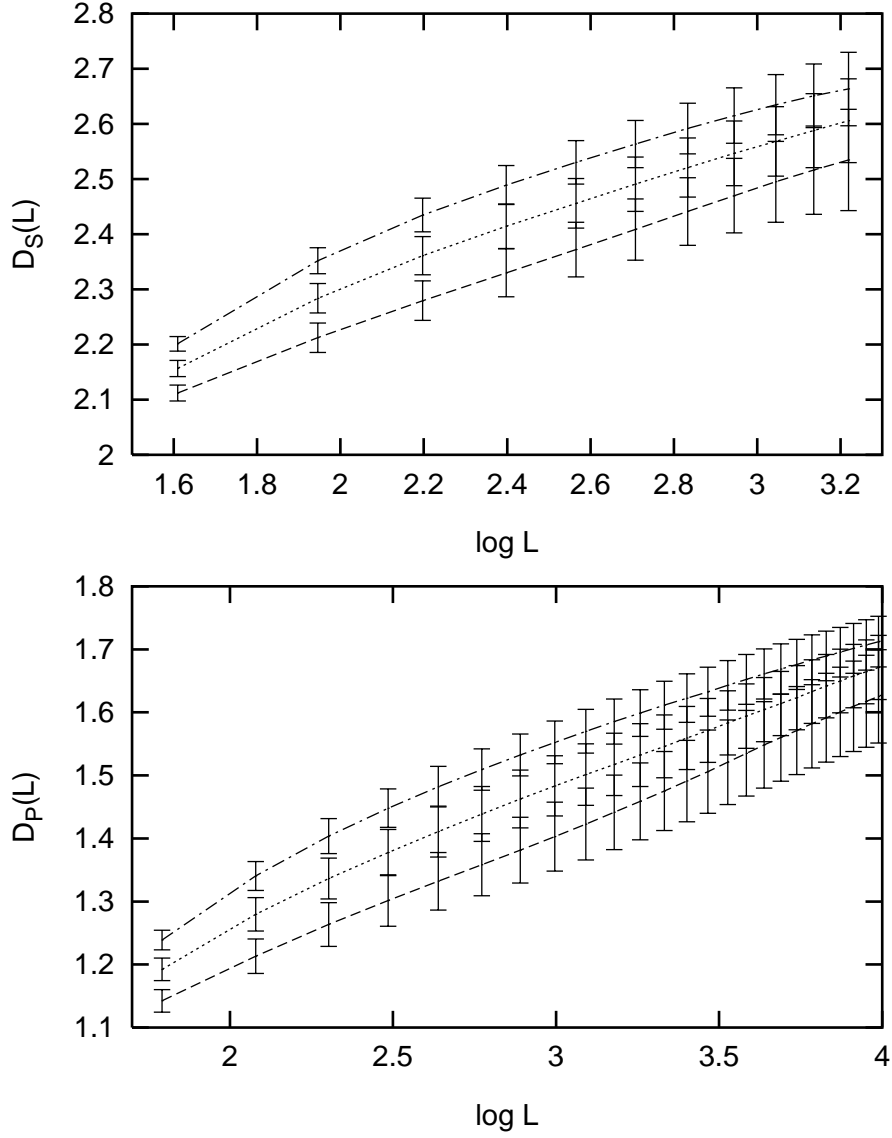}
\caption{The mean slope $D_S(L)$ and $D_P(L)$ and the dispersion
for noiseless simulations with $n=0.3$ (bottom), $n=1.$ (middle)
and $n=1.7$ (top).}
\label{dl}
\end{figure}

We have performed the surface and profile analysis on 
Monte Carlo simulations of the {\it COBE}--DMR maps for different 
values of the spectral index $n$. We used 500 simulations
of the maps for each value of $n$ from 0.3 to 1.7 (in steps
of 0.1). Even if we neglect the radiometer noise, we do not
expect a unique result for different realizations with the 
same spectral index due to the cosmic variance. The 
distribution of the results for the 500 simulation with a
given $n$ is a measure of the results that observers located
at different positions would obtain.
The mean curves of the slopes as a function of the scales over the 500
simulations for the surface analysis, $D_S(L)$, and for the 
profile analysis, $D_P(L)$, show a clear ordering according to 
their spectral index, as can be seen in Figure \ref{dl}.
 In both cases, the larger $n$ have larger
slopes associated at all scales. This result reflects the fact 
that a larger spectral index  gives rise to a rougher surface
as it is clearly illustrated by the comparison of
 the spheres corresponding to 
$n=0.3$ and $n=1.7$ in Figure \ref{spheres} (Plate 1)
\placefigure{dl}

This property
can be used to infer the most probable value of $n$ associated
to a (noiseless) map. For each of the simulated maps, we 
construct a six--dimensional vector $d_i$, with the first three 
components given by the $D_S(L)$ values at
three different scales and the last three components by the 
$D_P(L)$ values at three different scales. In this way, combining
the information from the surface and profile analysis, we 
make full use of the signal in the cut and uncut faces. The choice
of the three scales for each set is enough, as each point contains 
information from five consecutive points in the $V(L)$ or $S(L)$
curves. The inclusion of more points does not diminish the 
resulting dispersions.
The $\chi^2$ statistics is then obtained as
\begin{equation}
\chi^2=\sum_{i=1}^6 \sum_{j=1}^6 \left(\langle d_i
\rangle -d_i^M \right) M_{ij}^{-1} \left(\langle d_j
\rangle -d_j^M \right),
\label{chi2}
\end{equation}
where $\langle d_i \rangle$ is the mean of the $d_i$
computed from the 500 simulations of the maps for a given $n$
and the covariance matrix is given by
\begin{equation}
M_{ij}=\frac{1}{N}\sum_{k=1}^N \left(d_i^k-\langle d_i
\rangle \right) \left(d_j^k-\langle d_j
\rangle \right),
\label{covm}
\end{equation}
with $N=500$, the number of realizations used. We denoted by 
$d_i^M$ the slopes obtained for the map that we want to analyze.
The likelihood statistics ${\cal L}$ is given by
\begin{equation}
{\cal L}=\frac{1}{(2\pi)^{m/2} (\det M)^{1/2}} \exp(-\chi^2/2),
\label{lik}
\end{equation}
with $m=6$, the number of data points used in the fit.

We tested the method by using as input data 100 simulations
with $n=1$. We first computed the $d_i^M$ for each of them 
and then  obtained $\chi^2$ for the different $n$ values. 
We fitted a parabola  and associated the value of $n$
at the minimum as the best-fitting $n$ to the data. The returned
values of $n$ for the 100 simulations have mean 
$\langle n\rangle =1.01$ and dispersion $\sigma _n=0.16$.
A similar analysis using the maxima of the likelihood returned
$\langle n\rangle =1.02$ and $\sigma _n=0.19$.
Thus, both methods work reasonably well to determine $n$, the
dispersion measures the cosmic variance. As noted before,
this determination is independent of the amplitude of the spectrum 
(or the value of $Q$).

When analyzing actual sky maps, we have to take into account
the radiometer noise. For maps with a high signal to noise ratio
we expect that an analysis following the line presented in
this section will lead to a rather good determination of $n$.
However, when dealing with low signal to noise  maps, like 
the DMR one, this method is not useful anymore.
By one side, the ordering of the curves 
$D_{S(P)}(L)$ with respect to the $n$ value is not
maintained at all scales, but they cross in the small
scales region, where the noise is more important. Moreover, the 
method is no more independent of $Q$, the reason being 
that the amplitude of the noise introduces an intrinsic scale.
Hence, a rescaling by a constant factor of the sky signal does
not change the map by a constant factor, and the slope curves 
are no more independent of $Q$. 
We can however use a modified strategy to obtain the
$Q$ and $n$ best-fitting values.

\section{Analysis of the DMR  data}

In order to obtain the simultaneous determination of 
$Q$ and $n$ it is more useful to work directly 
with the volume of  the coverings of the surface  
as a function of the scale, $V(L)$, instead of  
the slopes, $D_{S(P)}(L)$, as the former 
are more sensitive to the amplitude of the spectrum (or
$Q$). In fact, in the noiseless case it was useful to 
work with the slopes because they provide a measure of the
roughness that is independent of the amplitude. As we are now
interested in determining also the amplitude, it is convenient
to go a step back and work with the volume of the coverings.

\begin{figure}
\epsfbox{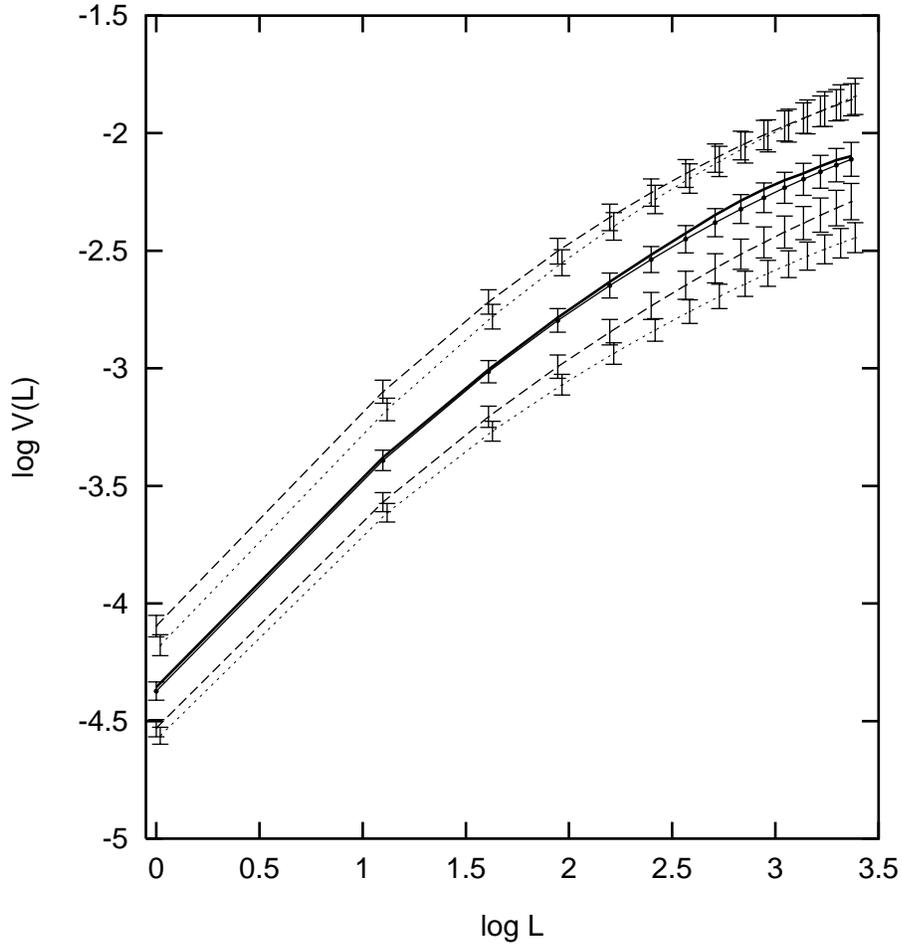}
\caption{$\log V$--$\log L$ plot for the {\it COBE} data (thick 
solid line), and the mean
curve and dispersion for 500 simulations with $Q=14 \mu$K and 
$n=1.2$, the best fit model (thin solid line), and  $Q=14 \mu$K, 
with $n=0.6$ (dashed bottom line) and  $n=1.8$ (dashed top line), 
and $n=1.2$, with $Q=8 
\mu$K (dotted bottom line) and $Q=20  \mu$K (dotted top line). }
\label{vol}
\end{figure}

We proceed in a way closely related to
the analysis of the surface in the last section. 
In this case, in order to have a smoothly varying field and 
reduce the effect of the radiometer noise, we applied a
Gaussian smoothing filter of $7^\circ$ (full width at
half maximum) to the maps before starting the analysis.
For every
map that we analyze we obtain the volume of the covering of
the surface at eight different scales ranging from 2 to 28 
pixels. To deal with the 
border problems, we consider the surface formed by the two
uncut faces, while we use the data in the cut faces to
compute the minima and maxima as in the last section.
The map used for our analysis is a weighted combination of 
the 53 and 90 GHz maps with the custom Galaxy cut 
(including the Orion and Ophiuco regions) of the 
4 yr {\it COBE}--DMR data (Bennett 
et al. 1996), as released by the NSSDC.
We simulate 500 maps following the characteristics of the map 
just mentioned, including the corresponding 
noise of the DMR combination for each individual pixel. 
We use for the multipole amplitudes the power-law  expression by
\cite{bo87,fa87}, that is a very good approximation at large scales for
a matter dominated universe (this can be readily generalized and can
consider the case of a cosmological constant or curvature dominated
universe using the multipoles obtained with the CMBFAST code
(\cite{se96})).
We consider 
values of $Q$ between 5 and 30 $\mu$K (in steps of
1 $\mu$K) and values of $n$ between $0$ and $2.2$ (in steps of 0.1). 
The curves $\log V(L)$ vs $\log L$ show an ordering,
with larger $V$ associated to larger $Q$ for a
fixed $n$, and to larger $n$ for a fixed $Q$.
We used these curves to test which is the model that best fits
the data.
In figure \ref{vol} we compare the curves $\log V(L)$  for 
several models with the one for the data.

The $\chi^2$ statistics is in this case given by
\begin{equation}
\chi^2=\sum_{i=1}^8 \sum_{j=1}^8 \left(\langle \log V_i
\rangle -\log V_i^M \right) M_{ij}^{-1} \left(\langle 
\log V_j\rangle -\log V_j^M \right),
\label{chi2v}
\end{equation}
where $\langle  \rangle$ denotes the mean over
the 500 simulations of the maps for a given $Q$ and $n$,
and the covariance matrix is given by
\begin{equation}
M_{ij}=\frac{1}{N}\sum_{k=1}^N \left(\log V_i^k-\langle 
\log V_i \rangle \right) \left(\log V_j^k-\langle 
\log V_j \rangle \right).\label{covmv}
\end{equation}
The likelihood is defined as in eq. (\ref{lik}), with $m=8$.
We consider here more scales than in the analysis
of the slopes performed in Section 3 because there the 
local dimension, being the  slope at a given scale,
contains information from all scales within the considered 
sliding window.
For a given map, we identify as the most probable values of
$Q$ and $n$ those which maximize ${\cal L}$.
The choice of eight 
points uses information from as many scales as possible 
without suffering numerical inaccuracies from inverting a matrix
that becomes very degenerate since the information from different 
scales is strongly correlated. 

We tested the accuracy of this method by applying it to 
simulations of the DMR maps as input. We used 900 simulations with 
$Q_{in}=18 \mu K$ and $n_{in}= 1$ and we obtained for each of them 
the values $Q_{ML}$ and $n_{ML}$ that maximize the likelihood.
The returned values for $Q_{ML}$ have a mean $\langle Q_{ML}
\rangle= 18.4 \mu K$ and a dispersion $\sigma_Q=4.4 \mu K$, while 
the returned values for $n_{ML}$ have a mean $\langle n_{ML}
\rangle= 0.99$ and a dispersion $\sigma_n= 0.33$.
Thus, the bias is very small both for the spectral index as well 
as for the amplitude estimation.

\begin{figure}
 \epsfbox{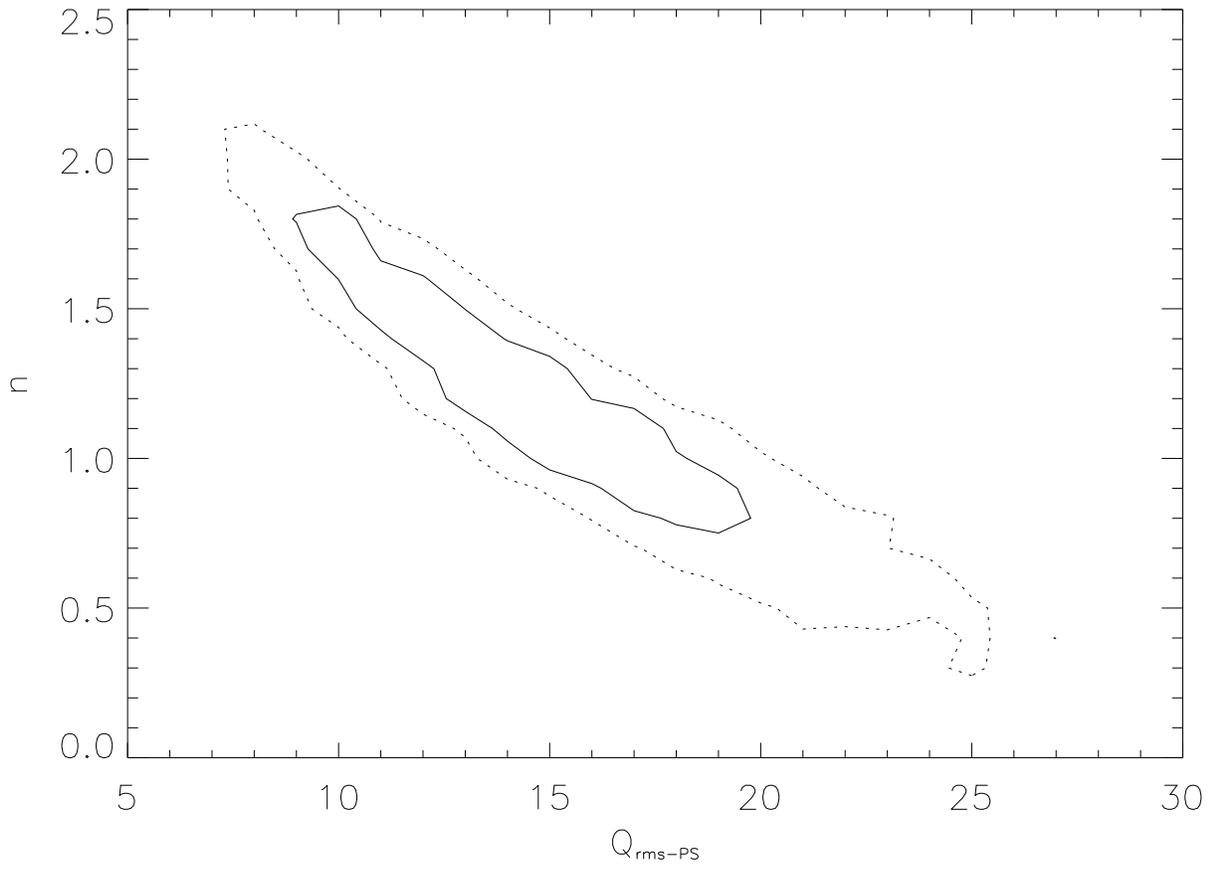}
\caption{Contour plot for the likelihood function  ${\cal L}(Q,n)$
corresponding to the $68 \%$ (solid line) and $95 \%$ (dotted line)
confidence levels.}
\label{cont}
\end{figure}

The analysis of the 4 yr DMR data with this method returns as the 
best-fitting values $Q= 14.1^{+3.9}_{-3.5}\mu K$ and  $n=
1.15^{+0.39}_{-0.34}$, where we have assigned the $68 \%$ credible 
interval to $Q$ and $n$ by marginalizing the two--dimensional 
likelihood ${\cal L}(Q,n)$. The 
likelihood function shows a strong anticorrelation between our 
estimates  of $Q$ and $n$. This can clearly be seen in Figure
\ref{cont}, which shows the contour plots corresponding to the 
$68 \%$ and $95 \%$ confidence regions of the likelihood function 
in the $Q - n$ plane. The contours show a significant elongation
along a line that can be parameterized by $Q(n)=32. \times 
\exp(-0.69 n)$.
These results are consistent with a scale invariant power spectrum, 
in this case the best-fitting normalization is given by 
$Q_{rms-PS}|_{n=1}= 16.2^{+1.4}_{-1.3}\mu K$.
These estimates are in agreement with the results of the {\it COBE}
group analysis of the data.  
\cite{be96} summarize their 4 yr results for the spectrum and 
normalization as $n=1.2\pm 0.3$ and $Q=15.3^{+3.8}_{-2.8} \mu$K, 
with the various analysis performed leading to very close values.
The determinations of $n$ and $Q$ presented in this paper have both
a central value slightly below the {\it COBE} ones, but well within the
1$\sigma$ range. For the scale invariant spectrum $n=1$, our 
normalization is 1$\sigma$ below the central value reported by the
COBE group, $Q|_{n=1}=18.\pm 1.6\mu$K. 
The error bars are comparable.
Results consistent with  those presented
in this paper are obtained by a multifractal analysis of the DMR
data (\cite{di99}).

We would like to note that in the previous analysis we have only
considered the effect of scalar perturbations on the CMB anisotropies. 
However, in many models of inflation leading to $n<1$ spectra, 
primordial gravitational waves also give a relevant contribution to 
large scale anisotropies (see e.g. \cite{lu92}). 
This leads to a decrement of the value of $Q$ associated to 
a given map (\cite{cr93}). 
The gravitational wave contribution is negligible for 
most inflationary models leading to $n>1$ (\cite{mo93}), and thus the 
amplitude determination is not affected in this case.

\section{Conclusions}

We have proposed a new method to analyze temperature
anisotropy maps that is based on the study of the roughness 
of a surface which has
heights and depths proportional to the temperature sky 
anisotropies. For large angle 
anisotropies produced by the Sachs--Wolfe effect such surface
should correspond to a fractal
surface, with fractal dimension $D$ directly related to the 
spectral index $n$ of the primordial fluctuations.
Thus, the determination of $D$ with known techniques developed 
for that purpose in other fields of science 
would lead to a determination of $n$. Unfortunately,
we found that the range of angular scales over which the Sachs--Wolfe 
effect dominates is not large enough to allow a sampling of the surface 
in enough points for obtaining a direct determination of $D$ as
the constant value of the plateau in the $D(L)$ plot. However, this does
not preclude a determination of $n$ for large signal to
noise anisotropy maps. We showed that in this case the whole 
(non-constant) $D(L)$ curve depends strongly on the spectral 
index and thus provide a useful tool to measure $n$. We tested 
this method with noiseless simulated maps and showed that it
returns an accurate determination of the spectral index. All 
this procedure is independent of the overall normalization 
of the map.

For low signal to noise maps, the previous method is no longer
useful to determine $n$ because the $D(L)$ curves for different 
$n$ cross at
small scales where the noise is more relevant. Furthermore,
the relative normalization of the signal to the amplitude 
of the noise also determines the shape of the $D(L)$ curve
(it is no longer function of $n$ alone). To overcome this problem,
we proposed an alternative method, in which the curve $V(L)$, 
corresponding to the volume of the surface covering vs. scale, 
is used instead of $D(L)$. This procedure allows a joint 
determination of $Q$ and $n$ in this case. 
The tests with simulated data showed that the estimated 
values are unbiased, and have reasonable error bars. 
The results of the 
application of this method to the {\it COBE}--DMR data, 
assuming a flat $\Omega=1$ cosmological model,  are 
fully compatible with those of more standard analysis.

Let us note that although the original motivation was to determine 
the (constant) spectral index at large scales measuring the 
fractal dimension $D$, the methods developed in Sections 3 and 4,
based on the study of the curves $D(L)$ and $V(L)$, are suitable
for comparing anisotropy maps to models leading to any kind of spectrum.
In particular, they can be used to analyze small angular resolution
maps, where a rich structure is expected to be found at small scales. 
In fact, 
the methods proposed here are just a different way to measure the
roughness (or structure) appearing in the maps as the scale changes.
Let us note that this kind of analyses can be performed for models
for which extensive Monte Carlo simulations over the range of
parameters of interest can be run. Thus, for the moment it is
difficult to apply it to the interesting case of topological defect
models.  Finally, we would like to stress the simplicity 
of the method from the computational point of view. To analyze a 
given map, we just need to compute at each scale the maximum and 
minimum of the $\Delta T$ in a neighborhood of every point. Using the 
cascade implementation of \cite{du89b}, at every scale we can use
the results of the previous scale, so the computational time of 
the algorithm is $O(N)$, with $N$ being the number of pixels.

\acknowledgements
We would like to thank R. B. Barreiro for kindly providing
her program for the simulations, L. Cay\'on for help 
dealing with the {\it COBE}--DMR maps and interesting discussions
and J. Schmalzing for comments on the manuscript.
SM acknowledges the Vicerrectorado de Investigaci\'on from 
U. de Valencia for a visiting professor position and CONICET for 
financial support. JMD thanks the DGES for 
a fellowship.  This work has been financially supported by  
the  Spanish DGES,  project n. PB95-1132-C02-02 and project 
n. PB96-0797, and by the Spanish CICYT,  project n. ESP96-2798-E. 
The COBE data sets were developed by the NASA Goddard Space 
Flight Center under the guidance of the {\it COBE}  Science Working 
Group and were provided by the NSSDC.

\newpage

\end{document}